\documentstyle[11pt, aaspp4]{article} 

\newcommand{\Teff} {T_{\rm eff}}
\newcommand{\logg}{\log g}
\newcommand{\vsini}{v \sin i}
\newcommand{\kms}{\, {\rm km} \, {\rm s}^{-1}}
\newcommand{\Weq}{W_\lambda}
\newcommand{\etal}{{\it et al.}}

\newcommand{\unit}[1]{\, {\rm #1}}

\lefthead{Behr}
\righthead{BHB Abundances and Rotations in M15}

\begin{document}

\title{Rotations and Abundances \\ of Blue Horizontal-Branch Stars \\ in Globular Cluster M15\altaffilmark{1}}

\author{Bradford B. Behr\altaffilmark{2}, Judith G. Cohen\altaffilmark{2}, \& James K. McCarthy\altaffilmark{3}}

\altaffiltext{1}{Based in large part on observations obtained at the
	W.M. Keck Observatory, which is operated jointly by the California 
	Institute of Technology and the University of California}
\altaffiltext{2}{Palomar Observatory, Mail Stop 105-24,
	California Institute of Technology, Pasadena, CA, 91125}
\altaffiltext{3}{PixelVision, Inc., 4952 Warner Avenue, Suite 300, Huntington Beach, CA, 92649}


\clearpage

\begin{abstract}

High-resolution optical spectra of eighteen blue horizontal-branch (BHB) stars in the globular
cluster M15 indicate that their stellar rotation rates and photospheric compositions vary strongly
as a function of effective temperature. Among the cooler stars in the sample, at $\Teff \sim 8500
\unit{K}$, metal abundances are in rough agreement with the canonical cluster metallicity, and the
$\vsini$ rotations appear to have a bimodal distribution, with eight stars at $\vsini < 15 \kms$ and
two stars at $\vsini \sim 35 \kms$. Most of the stars at $\Teff \ge 10000 \unit{K}$, however, are
slowly rotating, $\vsini < 7 \kms$, and their iron and titanium are enhanced by a factor of 300 to
solar abundance levels. Magnesium maintains a nearly constant abundance over the entire range of
$\Teff$, and helium is depleted by factors of 10 to 30 in three of the hotter stars. Diffusion
effects in the stellar atmospheres are the most likely explanation for these large differences in
composition. Our results are qualitatively very similar to those previously reported for M13 and
NGC~6752, but with even larger enhancement amplitudes, presumably due to the increased efficiency of
radiative levitation at lower intrinsic [Fe/H]. We also see evidence for faster stellar rotation
explicitly preventing the onset of the diffusion mechanisms among a subset of the hotter stars.

\end{abstract}



\clearpage

\section{Introduction}

M15 (NGC 7078) is one of the most metal-poor globulars known, with a metallicity ${\rm [Fe/H]} \sim
-2.4$~dex measured from red giant abundances (Cohen 1979, Sneden \etal\ 1997). Like many other such clusters, M15's horizontal
branch lies predominantly bluewards of the instability strip, and color-magnitude diagrams (Buonanno
\etal\ 1983; Durrell \& Harris 1993) show an extended ``blue tail'' reaching $\Teff$ as high as
$20000 \unit{K}$, which is separated from the horizontal part of the HB by a ``gap'' in the
distribution of stars along the HB (Moehler \etal\ 1995). Similar gaps appear in the CMDs of M13,
M80, NGC~6752, NGC~288, and other clusters, but are difficult to explain via standard models of RGB
mass loss or HB evolution.

Recently, attention has focused on atmospheric effects as a possible explanation for these
photometric features (Caloi 1999; Grundahl \etal\ 1998). At $\Teff \sim 10000 \unit{K}$, they
suggest, the stellar atmospheres become susceptible to diffusion effects, and thus develop chemical
peculiarities similar to those that appear in main-sequence CP stars. The resulting changes in
atmospheric opacity alter the emitted flux distributions and thus the measured photometry of the
hotter stars, giving rise to the gaps. These claims have been bolstered by measurements of large
photospheric abundance anomalies among hotter BHB stars in M13 (Behr \etal\ 1999) and NGC~6752
(Moehler \etal\ 1999), which show 30 to 50 times the iron abundance expected for these metal-poor
clusters. In M13, the transition from normal-metallicity cooler stars to metal-enhanced hotter stars
is remarkably abrupt, and coincides with the location of the gap, further supporting the
surface-effect hypothesis.

It appears, however, that stellar rotation also plays some role in this process. Theoretical
treatments of the diffusion mechanisms (Michaud \etal\ 1983) suggest that circulation currents
induced by higher rotation velocities can easily prevent abundance variations from appearing. This
prediction is borne out by measurements of $\vsini$ for the M13 stars (Behr \etal\ 2000), which show
that although the cooler stars exhibit a range of $\vsini$, some as high as $40 \kms$, the hot
metal-enhanced stars {\it all} show very low rotations, $\vsini < 10 \kms$. This correlation
suggests that slow rotation may be required in order for the metal enhancement and helium depletion
appear in the photosphere.

Although these results from M13 and NGC~6752 imply that we are on the right track towards
explaining the photometric peculiarities, BHB stars in many other clusters will have to be analyzed
in a similar fashion before we can make any firm claims. In particular, since the radiative
levitation that causes the observed metal enhancements is thought to depend strongly on the
intrinsic metallicity of the atmosphere, we should study clusters spanning a range of [Fe/H], to see
whether the onset and magnitude of the enhancements vary. To this end, we have observed BHB stars in
five clusters in addition to M13. In this Letter, we describe the results for both rotation and
photospheric abundances in M15. The data for the other clusters --- M92, M68, NGC~288, and M3 --- are being
analyzed currently, and will be presented in a future publication.


\section{Observations and Reduction}

The eighteen stars in our sample were selected from Buonanno \etal\ (1983), and are listed in Table
1. We acquired supplementary Str\"omgren photometry of the target stars using the Palomar 60-inch
telescope, in order to better constrain the effective temeratures. The stars generally lie in the
cluster outskirts, where crowding and confusion are less of a problem than towards the core, and our CCD imaging confirmed
an absence of faint companions.

The spectra were collected using the HIRES spectrograph (Vogt \etal\ 1994) on the Keck I telescope,
during four observing runs on 1997 August 01--03, 1997 August 26--27, 1998 June 27, and 1999 August
14--17. A 0.86-arcsec slit width yielded $R = 45000$ ($v = 6.7 \kms$) per 3-pixel resolution
element. We limited frame exposure times to 1200 seconds, to minimize susceptibility to cosmic ray
accumulation, and then coadded four frames per star. $S/N$ ratios were on the order of $30-60$
per resolution element.

We used a suite of routines developed by J.K. McCarthy (1988) for the FIGARO data analysis package
(Shortridge 1988) to reduce the HIRES echellograms to 1-dimensional spectra. Frames were
bias-subtracted, flat-fielded against exposures of HIRES' internal quartz incandescent lamps
(thereby removing much of the blaze profile from each order), cleaned of cosmic ray hits, and
coadded. A thorium-argon arc lamp provided wavelength calibration. Sky background was negligible,
and 1-D spectra were extracted via simple pixel summation. A 10th-order polynomial fit to line-free
continuum regions completed the normalization of the spectrum to unity.


\section{Analysis}

The resulting spectra show a few to over 140 metal absorption lines each, with the hottest stars showing the
largest number of lines. Line broadening from
stellar rotation is evident in a few of the stars, but even in the most extreme cases, the line
profiles were close to Gaussian, so line equivalent widths ($\Weq$) were measured by least-square
fitting of Gaussian profiles to the data. Equivalent widths as small as 10 m\AA\ were measured
reliably, and errors in $\Weq$ (estimated from the fit $\chi^2$) were typically 5 m\AA\ or less.
Lines were then matched to the atomic line lists of Kurucz \& Bell (1995), and those that were
attributed to a single species ({\it i.e.} unblended) were used to determine radial velocity $v_r$
for each of the stars. On the basis of $v_r$, all of the targets appear to be cluster members.

To derive photospheric parameters $\Teff$ and $\logg$, we compared the published photometry and our
own Str\"omgren indices to synthesized colors from ATLAS9 (Kurucz 1997), adopting a cluster
reddening of $E(B-V) = 0.09$. Temperatures are well-determined ($\pm$ a few hundred K) for the
cooler stars, but are somewhat less firm (as much at $\pm 1000 \unit{K}$) for the hotter stars. This
situation will improve with a more sophisticated treatment of the various Str\"omgren colors, and
by using transitions with different excitation potentials $\chi$ to constrain $\Teff$. We estimated surface
gravities using the $AB_\nu$ flux method (Oke \& Gunn 1984), which relates $\log g$, stellar mass
$M_*$, distance $d$, and photospheric Eddington flux $H_\nu$ at 5480~\AA. Since $AB_\nu(5480 \AA) =
V$ magnitude, we can derive
$$
\log g = 3.68 + \log(M_*/M_\odot) + \log(H_\nu) - (M-m)_V + 0.4 V_0 \: ,
$$
where unextincted magnitude $V_0 = V + A_V = V + 0.30$ for M15. We assumed $M_* = 0.6 M_\odot$ as a
representative BHB star mass, used a distance modulus of 15.26 (Silbermann \& Smith 1995), and drew the $H_\nu$ 
values from the ATLAS9 model atmospheres (Kurucz 1997), iterating until $\logg$ converged. The resulting
$\logg$ values agree well with model ZAHB tracks (Dorman 1993), except for the hotter stars, which
are ``overluminous'' as described by Moehler \etal\ (1995). Table 1 lists the final photospheric
parameters used for each of the target stars, as well as the heliocentric radial velocities.

For the chemical abundance analyses, we use the LINFOR/LINFIT line formation analysis package,
developed at Kiel, based on earlier codes by Baschek, Traving, and Holweger (1966), with subsequent
modifications by M. Lemke. Our spectra of these very metal-poor stars are sufficiently uncrowded
that we can simply compute abundances from equivalent widths, instead of performing a full spectral
synthesis fit. Only lines attributed to a single chemical species were considered; potentially
blended lines are ignored in this analysis. Upper bounds on [Ti/H] were determined for two of the
hotter stars, which did not show any titanium lines, by assigning an equivalent width of 20~m\AA\
(double the strength of the weakest metal lines actually measured in those spectra) to the six strongest Ti II transitions
and calculating the implied abundance. Microturbulent velocity $\xi$ was chosen such that the
abundance derived for a single species (Fe II for most cases) was invariant with $\Weq$. For those
stars with an insufficient number of lines to utilize this technique, we adopted a typical value of
$\xi = 2 \kms$. We assumed a cluster metallicity of [Fe/H]~$= -2.4$~dex in computing the initial
model atmospheres, and then adjusted as necessary for those stars that turned out to be considerably
more metal-rich, although these adjustments to the atmospheric input were found to have only modest
effects ($< 0.2$ dex) on the abundances of individual elements.

Measurement of $\vsini$ broadening traditionally entails cross-correlation of the target spectrum
with a rotational reference star of similar spectral type, but this approach assumes that the
template star is truly at $\vsini = 0$, which is rare. Furthermore, given the abundance
peculiarities that many of these BHB stars exhibit, it is difficult to find an appropriate spectral
analog. Since we are able to resolve the line profiles of our stars, we instead chose to measure
$\vsini$ by fitting the profiles directly, taking into account other non-rotational broadening
mechanisms. We used bright unsaturated arc lines to construct an instrumental profile, and then
combined it with an estimated thermal Doppler broadening of $3 \kms$ FWHM and the
previously-determined $\xi$. This profile was convolved with hemicircular rotation profiles for
various $\vsini$ to create the final theoretical line profiles. Each observed line in a spectrum was
fit to the theoretical profile using an iterative least-squares algorithm, solving for $\vsini$ and
line depth. The values for $\vsini$ from different spectral lines generally agree quite well, once we 
removed helium lines and blended lines such as the Mg II 4481 triplet.


\section{Results}

In Figure 1, abundance determinations for key chemical species are plotted as a function of stellar $\Teff$.
The values [X/H] represent logarithmic offsets from the solar values of Anders \& Grevesse (1993).
The error bars incorporate the scatter among multiple lines of the same species, plus 
the uncertainties in $\Teff$, $\logg$, $\xi$, $\Weq$ for each line, and [Fe/H] of the input atmosphere.

In the cooler stars ($\Teff < 10000 \unit{K}$), the compositions are largely as expected. The iron abundance
[Fe/H] averages $-2.5$, slightly below the value of $-2.4$ expected for this metal-poor cluster. Magnesium and
titanium appear at [Mg/H]~$\sim -2.2$ and [Ti/H]~$\sim -1.8$, respectively, which are also reasonable
for this environment.

As we move to the hotter stars, however, the iron abundances change radically. Six of the eight stars at
$\Teff \ge 10000 \unit{K}$ show solar iron abundances, [Fe/H]~$\sim 0$, an enhancement of a factor of 300
from the cooler stars. (The other two hot stars show the same [Fe/H] as the cool stars, for reasons which will be discussed
shortly.) Titanium, in a similar fashion, rises to [Ti/H]~$> 0$, although there are hints
of a monatonic increase with $\Teff$ rather than an abrupt jump. Magnesium, on the other hand, appears to
be unaltered, maintaining a roughly constant metal-poor level across the entire temperature range. The hotter
stars also start to show helium lines, providing evidence of helium depletion, since [He/H]~$\sim 0$ at 
$11000 \unit{K}$ but it then drops by factors of 10 to 30 for $\Teff = 12000$--$13000 \unit{K}$.

Figure 1 also charts the values of $\vsini$ derived for the target stars. Among the cooler stars, we find a range of 
rates, with most of the stars rotating at $15 \kms$ or less, except for two stars at $29 \kms$ and $36 \kms$, respectively.
This sort of distribution of $\vsini$ is not what one would expect given a single intrinsic rotation speed $v$ and
random orientation of the rotation axes, since large $\sin i$ are more likely than small $\sin i$.
Instead, the cool end of the HB appears to contain two rotational populations, one with $v \sim 35 \kms$, and another with $v \sim 15 \kms$,
much like in M13 (Peterson \etal\ 1995).

For the hotter stars, there also appears to be a bimodal distribution in $v$, although the difference is less pronounced.
Six of these eight stars show $\vsini < 7 \kms$, while the other two have $\vsini \sim 12 \kms$. Interestingly,
these two faster-rotating hot stars are the same stars that show ``normal'' (metal-poor) iron abundances.


\section{Discussion}

These BHB stars in M15 exhibit abundance and rotation characteristics very similar to those reported for
BHB stars in M13 (Behr \etal\ 1999 \& 2000). The enhancements of metals except for magnesium, the
depletion of helium, and the difference in maximum $\vsini$ between hotter and cooler stars are shared
by both clusters. The abundance anomalies in M15 are therefore likely to be due to the same diffusion processes 
that were invoked for the prior study --- radiative levitation of metals, and gravitational settling of helium,
in the stable non-convective atmospheres of the hotter, higher-gravity stars, as hypothesized by Michaud \etal\
Unfortunately, the stars selected in M15 do not sample the immediate vicinity of the photometric gap as
well as those in M13, so the association between the onset of diffusion-driven abundance variations and the
gap is not as clear-cut, but the general trend still supports this association.

There are, however, two notable differences between the results for M15 and those for M13. First, the
magnitudes of the metal enhancements are somewhat different. Iron and titanium are each enhanced by $\sim 2.5 \unit{dex}$ in
M15, while in M13, iron increases by only $\sim 2 \unit{dex}$, and titanium by $\sim 1.5 \unit{dex}$.
Despite this difference, the enhancement mechanism yields the same final metallicities for the hot stars in both clusters: [Fe/H]~$\sim 0$
and [Ti/H]~$\sim 0$. This correspondence suggests that the radiative levitation mechanism reaches equilibrium with gravity
at or near solar metallicity, independent of the initial metal content of the atmosphere, as the metal lines become saturated 
and are thus unable to support further enhancements.

Second, there is the issue of the two hotter M15 stars which do not show metal enhancement, denoted by circles in
Figure 1. Their derived temperatures associate
them with the hotter population, as does their photometry, which places them bluewards of the photometric gap. Their
iron abudances, however, are $< -2.5 \unit{dex}$, like the cooler unenhanced stars in the cluster, and stringent upper
bounds on their titanium abundances again suggest that they are metal-poor. These two stars
are also distinguished by having $\vsini \sim 12 \kms$, nearly twice as large as any of the other hot stars. Although
their temperatures and gravities are high enough to support radiative levitation, it appears that their faster rotations
induce meridonal circulation, which keeps the atmosphere well-mixed and prevents the metal enhancements
from appearing. Such sensitive dependence upon rotation speed was mentioned by Michaud \etal, but these
observations provide direct evidence that rotationally-driven mixing can and does directly influence the atmospheric composition.

With these results from M15, we add to the growing body of evidence that element diffusion, regulated by
stellar rotation, are at least partially responsible for the observed photometric morphology of globular cluster
HBs. The findings from this cluster corroborate the prior work on M13 and NGC~6752, while adding some
new twists which further illuminate the diffusion mechanisms. Analysis of many other clusters, spanning
a range of metallicity and HB morphology, will be necessary before all of the implications of these effects
can be known.


\acknowledgments

These observations would not have been feasible without the HIRES spectrograph and the Keck I telescope. We are indebted
to Jerry Nelson, Gerry Smith, Steve Vogt, and many others for making such marvelous machines, to the W. M. Keck
Foundation for making it happen, and to a bevy of Keck observing assistants for making them work. J.G.C. and B.B.B. acknowledge
support from NSF Grant AST-9819614.



\clearpage
 
\figcaption[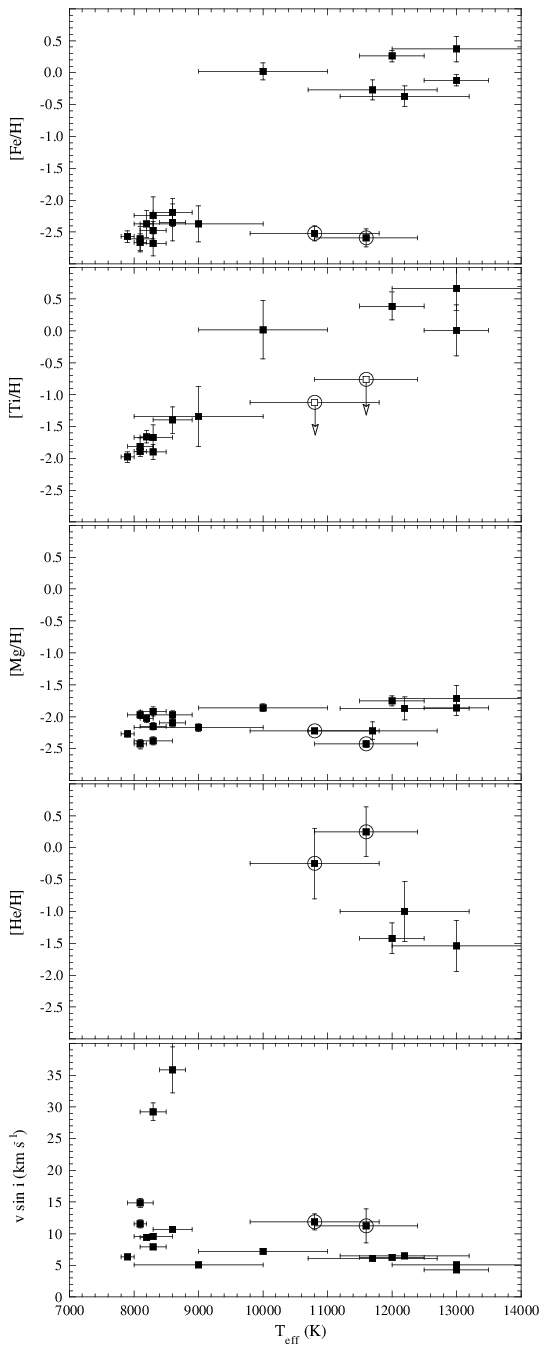]{Fe, Ti, Mg, and He abundances, and projected rotation velocity $\vsini$ for BHB
stars over a range of $\Teff$. Abundances are plotted as log offsets from the solar abundances.
Circles indicate the two hot fast-rotating non-metal-enhanced
stars discussed in the text, and the open symbols in the [Ti/H] plot are upper bounds on the titianium abundance
for these two stars.}

\epsscale{0.5}
\plotone{fig1.eps}


\clearpage 

\begin{deluxetable}{llrcrrccr}
\tablenum{1}
\tablewidth{0pt}
\scriptsize
\tablecaption{Parameters for program stars}
\label{tab1}
\tablehead{
			&			&		&				&$v_r$\quad ~&$\Teff$~\quad ~		&				&$\xi$ 		&$\vsini\;\;$~	\nl
Star			&\quad $V$	&$B-V$	&$N_{\rm lines}$	&($\kms$)	&(K)~\quad ~			&$\logg$			&($\kms$)	&($\kms$)
}
\startdata
B124 		&15.91		&0.15~		&47			&$-106.0$	&$7900 \pm\;\: 100$	&$2.80 \pm 0.05$	&2			&$6.38 \pm 0.25$		\nl
B558 		&15.93		&0.14~		&18			&$-95.3$		&$8100 \pm\;\: 100$	&$3.00 \pm 0.10$	&2			&$11.55 \pm 0.63$		\nl
B218 		&15.99		&0.16~		&16			&$-96.7$		&$8100 \pm\;\: 200$	&$2.95 \pm 0.05$	&2			&$14.88 \pm 0.69$		\nl
B78 			&15.99		&0.15~		&34			&$-110.8$	&$8200 \pm\;\: 100$	&$3.05 \pm 0.05$	&2			&$9.45 \pm 0.25$		\nl
B153 		&15.95		&0.14~		&7			&$-113.8$	&$8300 \pm\;\: 200$	&$2.95 \pm 0.05$	&2			&$29.25 \pm 1.35$		\nl
B244 		&15.96		&0.14~		&23			&$-116.0$	&$8300 \pm\;\: 300$	&$3.05 \pm 0.10$	&1			&$9.59 \pm 0.41$		\nl
B331 		&16.04		&0.14~		&12			&$-108.4$	&$8300 \pm\;\: 200$	&$3.05 \pm 0.05$	&2			&$7.93 \pm 0.34$		\nl
B177 		&16.03		&0.15~		&25			&$-110.6$	&$8600 \pm\;\: 300$	&$3.00 \pm 0.05$	&2			&$10.70 \pm 0.37$		\nl
B424 		&15.89		&0.14~		&6			&$-106.5$	&$8600 \pm\;\: 200$	&$3.05 \pm 0.10$	&2			&$35.81 \pm 3.63$		\nl
B130 		&15.96		&0.15~		&22			&$-115.2$	&$9000 \pm 1000$		&$3.05 \pm 0.10$	&2			&$5.07 \pm 0.24$		\nl
B267		&16.72		&0.03~		&109		&$-115.2$	&$10000 \pm\;\: 100$	&$3.55 \pm 0.10$	&0			&$7.22 \pm 0.17$		\nl
B334		&16.58		&0.02~		&8			&$-109.5$	&$10800 \pm 1000$	&$3.55 \pm 0.10$	&2			&$11.86 \pm 1.25$		\nl
B348		&16.69		&0.01~		&7			&$-107.7$	&$11600 \pm\;\: 800$	&$3.60 \pm 0.10$	&2			&$11.20 \pm 2.68$		\nl
B84			&16.56		&0.00~		&49			&$-109.0$	&$11700 \pm 1000$	&$3.60 \pm 0.10$	&0			&$6.12 \pm 0.20$		\nl
B279		&16.56		&0.01~		&144		&$-104.8$	&$12000 \pm 1000$	&$3.60 \pm 0.10$	&0			&$6.25 \pm 0.11$		\nl
B203		&16.68		&$-0.01$~	&57			&$-95.2$		&$12200 \pm 1000$	&$3.60 \pm 0.10$	&0			&$6.47 \pm 0.27$ 		\nl
B315		&16.80		&$-0.02$~	&48			&$-104.6$	&$12800 \pm\;\: 800$	&$3.75 \pm 0.10$	&0			&$4.26 \pm 0.18$		\nl
B374		&16.79		&$-0.02$~	&91			&$-107.2	$	&$13000 \pm 1000$	&$3.70 \pm 0.10$	&0			&$5.10 \pm 0.16$		\nl
\nl
\multicolumn{4}{l}{mean and dispersion}					&\multicolumn{2}{l}{\quad$-106.6 \pm 7.0$}
\enddata
\end{deluxetable}


\end{document}